\documentclass{article}
\usepackage{eurosym}
\usepackage{amsmath,amssymb,amsthm}
\usepackage{graphicx}
\usepackage{subcaption}
\usepackage{verbatim}
\usepackage{hyperref}
\usepackage{bm}
\usepackage{color}
\usepackage[numbers]{natbib}
\usepackage{amsmath}
\usepackage{amsfonts}
\usepackage{amssymb}

\begin{document}

\title{Impedance Matching in an Elastic Actuator}
\author{Tianyi Guo$^{1}$ \thanks{%
Email: tguo2@kent.edu}, Xiaoyu Zheng$^{2}$\thanks{
Email: xzheng3@kent.edu}, Peter Palffy-Muhoray$^{1,2}$\thanks{%
Corresponding author. Email: mpalffy@kent.edu} \\
\emph{$^1$Advanced Materials and Liquid Crystal Institute, Kent State
University, OH, USA}\\
\emph{$^2$Department of Mathematical Sciences, Kent State University, OH, USA%
} }
\maketitle

\begin{abstract}
We optimize the performance of an elastic actuator consisting of an active core in a host which performs mechanical work on a load. The system, initially with localized elastic energy in the active component, relaxes and distributes energy to the rest of the system.  Using the linearized Mooney-Rivlin hyperelastic model in a cylindrical geometry and assuming the system to be overdamped,  we show that the value of the Young's modulus of the impedance matching host which maximizes the energy transfer from the active component to the load is the geometric mean of Young's moduli of the active component and the elastic load. This is similar to the classic results for impedance matching for maximizing the transmittance of light propagating through dielectric media.
\end{abstract}


\section{Introduction}

When light propagates through a planar interface between two perfect
dielectrics, a portion of the light is reflected and the rest is
transmitted. To minimize the reflectance in medium 1 with refractive index $%
n_{1}$, or equivalently, to maximize the transmitted light to medium 3 with
refractive index $n_{3}$, one can insert an index matching layer with
refractive index $n_{2}=\sqrt{n_{1}n_{3}}$ between the two media.
Furthermore, the reflectance is zero if the thickness of the index matching
medium is one quarter of the wavelength \cite{Stratton}. One can apply the
same principle to achieve the perfect sound transmittance by positioning a
quarter wavelength impedance matching layer with index $\sqrt{(\rho
_{1}c_{1})(\rho _{3}c_{3})}$, with $\rho _{i}$ the mass density and $c_{i}$
the speed of sound \cite{Kim}. Impedance matching techniques are widely used
in applications involving elastic wave propagation as well as in
electronics \cite{Chen2014, Armor, Rathod2019}.

The case of a head-on elastic collision of two rigid balls with masses $%
m_{1} $ and $m_{3}$ offers an interesting analogy. To maximize the energy
transfer from $m_{1}$, which has nonzero initial energy $E_{1}$, to $m_{3}$,
which has zero initial energy, one can position a rigid ball with mass $%
m_{2}=\sqrt{m_{1}m_{3}}$ and zero initial energy, in between the two balls 
\cite{Collision}.

The similarity of these very different physical phenomena is that energy is
conserved throughout the process: energy is either reflected or transmitted.
The transmitted energy can be increased when an impedance matching medium is
inserted. The fraction of reflected energy in case of normal
incidence/collision between two media/balls, is given by, 
\begin{equation}
R_{12}=\left( \frac{Z_{2}-Z_{1}}{Z_{2}+Z_{1}}\right) ^{2},  \label{eq_R}
\end{equation}%
where $Z_{i}=n_{i}$ for the case of light propagation through an interface,
and $Z_{i}=m_{i}$ for elastic collision of rigid balls. The transmitted
energy is given by $T_{12}=1-R_{12}$. If there are three media in series,
the fraction of transmitted energy from medium $1$ to medium $3$, via medium 
$2$, is given by 
\begin{equation}
T_{13}=T_{12}T_{23}.  \label{eq_T}
\end{equation}%
Here the interference due to multiple reflections has been neglected in Eq.\ \eqref{eq_T}, and
media $1$ and $3$ are assumed to be semi-infinite in the propagation
direction. Exact expression including the dependence on the thickness of layer $2$ in the wave
propagation case can be found in Refs.\ \cite{Stratton, Kim}. Upon maximizing $%
T_{13}$ in Eq.\ \eqref{eq_T} with respect to $Z_{2}$, one arrives
immediately at, 
\begin{equation}
Z_{2}=\sqrt{Z_{1}Z_{3}},  \label{eq_Z}
\end{equation}%
thus the value of $Z_{2}$ which maximizes the energy transmission from $1$
to $3$ is the geometric mean of $Z_{1}$ and $Z_{3}$. Mechanical impedance is
a measure of effectiveness of a force in producing velocity. Remarkably, one
can optimize certain energy transfer processes by inserting an index
matching component.

In this paper, we study a related problem of optimizing the transfer of
elastic energy from one elastic body to another via an impedance matching
element.

\begin{figure}[th]
\centering
\includegraphics[width=.5\textwidth]{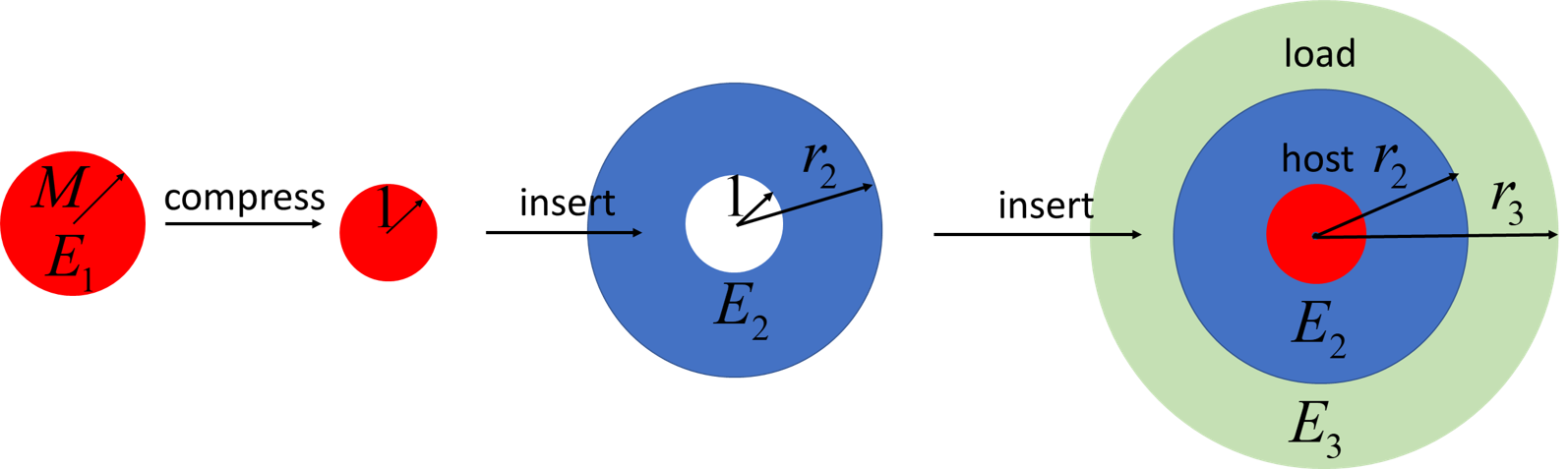}
\caption{A schematic showing the three bodies, the inner disk and two
annuli, each with different elastic modulus.}
\label{fig-annulus}
\end{figure}

Specifically, we consider three elastic bodies: body $1$ is the active
element with stored elastic energy, contained in body $2$, the host, which
is the impedance matching element and body $3$ is the load to which we wish
to transfer elastic energy. For simplicity, we use cylindrical symmetry in
our example. All three bodies are isotropic, homogeneous and uniform and
share the same axis of symmetry; body $1$ is a cylinder, while bodies $2$
and $3$ are annuli. The geometry is shown in Fig. \ref{fig-annulus}.

We begin with the host, body $2$, which is an annulus with a cylindrical
cavity. Initially it is stress free. We then take another elastic body, with
a different elastic modulus, which is too large (or too small) to fit fully
into the cavity of the host. We then compress (or stretch) this body until
its shape is the same as that of the cavity. This is the active body $1$. We
then place the active body, keeping its shape fixed, into the cavity of the
host. The third body is the load; an annulus whose cavity can perfectly
accommodate the host. Finally, we place the host with the active body into
the cavity of the load, as indicated in Fig.\ \ref{fig-annulus}. The system
is then allowed to relax.

When released, the internal stored elastic energy of the active medium will
do mechanical work on the load. The situation illustrated here is similar to
a light driven actuator, where the photoactive part of the system expands or
shrinks on illumination, distributing stress to the surrounding medium,
causing a deformation. \ The system can then do mechanical work, say expand
against a pressure. Given the properties of the actuator and the load, can
we maximize the work by choosing a suitable host material? Below, we present
a mathematical model of the deformation of elastic media in a cylindrical
geometry, and determine Young's modulus of the impedance matching host which
maximizes the energy transferred to the load. The results suggest a strategy
for optimizing the performance of an elastic actuator.

We remark that the similar analysis cannot be carried out in a spherical
geometry with volume conserving materials, since an incompressible sphere
cannot be radially deformed.

\section{Mathematical Model}

We use Lagrangian mechanics to model the system. We start with the
incompressible Mooney-Rivlin's hyperelastic model, in which the energy
density of an elastic material is a linear combination of invariants of the
left Cauchy-Green deformation tensor \cite{Mooney, Rivlin}%
\begin{equation}
W=C_{1}(\lambda _{1}^{2}+\lambda _{2}^{2}+\lambda _{3}^{2}-3)+C_{2}(\lambda
_{1}^{2}\lambda _{2}^{2}+\lambda _{1}^{2}\lambda _{3}^{2}+\lambda
_{2}^{2}\lambda _{3}^{2}-3),
\end{equation}%
where $\lambda _{i},i=1,2,3$ are principal stretches, and $\lambda
_{3}=1/(\lambda _{1}\lambda _{2})$ due to incompressibility. Assuming the
deformations are small, expanding in terms of $\lambda _{1}-1$ and $\lambda
_{2}-1$, we get,%
\begin{equation}
W=4(C_{1}+C_{2})((\lambda _{1}-1)^{2}+(\lambda _{2}-1)^{2}+(\lambda
_{1}-1)(\lambda _{2}-1)).
\end{equation}%
We further assume that all deformations have cylindrical symmetry, and
denote the position of a point in body $R(r)\mathbf{\hat{r}}+Z(z)\mathbf{%
\hat{z}}$. Here $r$ and $z$ are the Lagrangian coordinates denoting the
position of mass points in the undeformed system. Then the principal
stretches are given by%
\begin{equation}
\lambda _{1}=\frac{\partial R}{\partial r},\lambda _{2}=\frac{R}{r},\lambda
_{3}=\frac{\partial Z}{\partial z}=\frac{1}{\lambda _{1}\lambda _{2}},
\end{equation}%
where $\lambda _{1}$ is along the radial, $\lambda _{2}$ along the azimuthal
and $\lambda _{3}$ along the $z-$ direction. In terms of $R(r)$, we can
express the elastic energy density in the linear regime as%
\begin{equation}
W_{E}=\frac{2}{3}E((R^{\prime }-1)^{2}+(\frac{R}{r}-1)^{2}+(R^{\prime }-1)(%
\frac{R}{r}-1)),
\end{equation}%
where $E=6(C_{1}+C_{2})$ is Young's modulus and $R^{\prime }=\partial
R/\partial r$.

The energy per length in the $z-$direction of the system consisting of the elastic bodies $1$, $2$ and 
$3\,\ $is given by%
\begin{equation}
F=2\pi \left(
\int_{0}^{M}W_{E_{1}}dr+\int_{1}^{r_{2}}W_{E_{2}}dr+%
\int_{r_{2}}^{r_{3}}W_{E_{3}}dr\right) ,
\end{equation}%
where the radius of the central hole in the undeformed host is taken to be
unity, $M$ is the radius of the pre-strained active core, $r_{2}$ and $r_{3}$
are the outer radii of the host and the load, respectively. 

Minimizing the total energy $F$ gives the Euler-Lagrange equation describing
the deformation. All three parts, active core, host, and load, share the
same form of the equation, which is given by%
\begin{equation}
R^{\prime \prime }r+R^{\prime }-\frac{R}{r}=0.
\end{equation}%
It admits the solution%
\begin{equation}
R_{i}(r)=A_{i}r+\frac{B_{i}}{r},
\end{equation}%
where $A_{i}$ and $B_{i}$, $i=1,2,3$, are determined by the interface and
boundary conditions, which are detailed below.

The continuity condition for displacements across the interfaces are given
by,%
\begin{eqnarray}
R_{1}(0)\text{ is finite, } &&\text{or }B_{1}=0, \\
R_{1}(M)=R_{2}(1), &&\text{or }A_{1}M=A_{2}+B_{2}, \\
R_{2}(r_{2})=R_{3}(r_{2}), &&\text{or, }A_{2}r_{2}+\frac{B_{2}}{r_{2}}%
=A_{3}r_{2}+\frac{B_{3}}{r_{2}}.
\end{eqnarray}%
In addition, the normal stresses are continuous across the two interfaces,
which are 
\begin{eqnarray}
E_{1}(2R_{1}^{\prime }+\frac{R_{1}}{r}-3)r|_{r=M} &=&E_{2}(2R_{2}^{\prime }+%
\frac{R_{2}}{r}-3)r|_{r=1}, \\
E_{2}(2R_{2}^{\prime }+\frac{R_{2}}{r}-3)r|_{r=r_{2}}
&=&E_{3}(2R_{3}^{\prime }+\frac{R_{3}}{r}-3)r|_{r=r_{2}}.
\end{eqnarray}%
We would need another boundary condition at the outmost boundary to complete
the set of equations to be solved. 

\subsection{Zero-strain boundary condition}

If the outer boundary of the load is fixed, boundary condition at $r_{3}$
reads as,%
\begin{equation}
R_{3}(r_{3})=r_{3}.
\end{equation}%
Together with the five interface conditions, these six linear equations
determine the six unknowns, $A_{i}$ and $B_{i}$ uniquely, and they are
functions of $E_{i}$ and $r_{i}$, $i=1,2,3$. Since the solutions are rather
lengthy algebraic expressions, we omit them and only report the final
optimization results.

We are interested in the transfer of elastic energy from the active core to
the outside load. We therefore ask: what value of Young's modulus $E_{2}$ of
the host material will maximizes the transfer of energy from the active core
to the load? Maximizing the energy in the load transferred from the active
core is equivalent to maximizing the displacement of inner radius of the
load $R_{3}(r_{2})$. We note that $R_{3}(r_{2})=A_{3}r_{2}+B_{3}/r_{2}$.
Taking the derivative of $R_{3}(r_{2})$ with respect to $E_{2}$, we obtain,%
\begin{equation}
E_{2}=\sqrt{E_{1}E_{3}}\sqrt{\frac{3r_{2}^{2}+r_{3}^{2}}{r_{3}^{2}-r_{2}^{2}}%
}.  \label{eq_index_zero_strain}
\end{equation}%
In the case of
$r_{3}\rightarrow \infty $, the load is infinitely large, we have $E_{2}=%
\sqrt{E_{1}E_{3}}$. This results is a reminiscent of an equivalent result for refractive indices in the case of impedance matching for light
propagation in 1D media. 

\subsection{Zero-stress boundary condition}

In this case, the outer boundary of the load is free to move, and the
boundary condition at $r_{3}$ is given by,%
\begin{equation}
(2R_{3}^{\prime }+\frac{R_{3}}{r}-3)|_{r=r_{3}}=0.
\end{equation}%
Together with the interface equations, the six unknowns, $A_{i}$ and $B_{i}$
are uniquely determined. Again, we are interested in the energy transfer
from the active core to the load, and we ask the same question as in the
zero-strain case: what value of Young's modulus $E_{2}$ of the host material
will maximize the transfer of energy from the active core to the load?
Although it is not obvious, maximizing the energy in the load transferred
from the active core is equivalent to maximizing the displacement of outer
radius of the load $R_{3}(r_{3})$. Taking the derivative of $R_{3}(r_{3})$
with respect to $E_{2}$, we obtain 
\begin{equation}
E_{2}=\sqrt{E_{1}E_{3}}\sqrt{\frac{3r_{3}^{2}-3r_{2}^{2}}{%
r_{2}^{2}+3r_{3}^{2}}}.  \label{eq_index_zero_stress}
\end{equation}%
Again the result is a geometric mean of Young's modulus of medium 1 and 3,
multiplied with a geometric factor depending on radii of components. In the
case when $r_{3}\rightarrow \infty $, $E_{2}=\sqrt{E_{1}E_{3}}.$

\begin{figure}[th]
\centering
(a)\includegraphics[width=.45\textwidth]{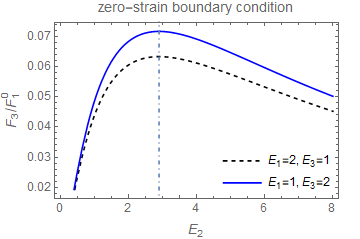}  (b)%
\includegraphics[width=.45\textwidth]{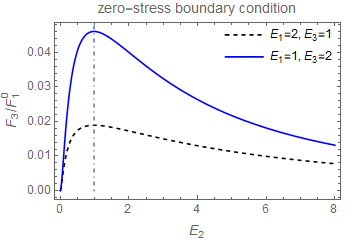}  
\caption{Transferred energy to the load $F_3$ normalized by initial energy of
active component $F_1^0$ as a function of $E_2$ with $M=2,r_2=2,r_3=3$. The vertical
dotdashed lines are located at the optimal value of $E_2$ (a.u.), given by
Eq.\ \eqref{eq_index_zero_strain} and Eq.\ \eqref{eq_index_zero_stress},
respectively. (a) zero-strain boundary condition, (b) zero-stress boundary
condition. }
\label{fig-fixed-free}
\end{figure}

Figure \ref{fig-fixed-free} demonstrates the maximum energy transferred to the load from the active material occurs when the impedance
matching modulus $E_{2}$ is given by Eq.\ \eqref{eq_index_zero_strain} or %
\eqref{eq_index_zero_stress} at zero-strain or zero-stress boundary
condition, respectively.

\subsection{Reflectance and transmittance}

We finally look at the problem from the reflectance and transmittance point
of view and build a connection with the case of light propagation. Consider
the active core and the load only, with fixed outer boundary condition. The
active core initially has stored elastic energy; it is subsequently released
and transfers some of its stored energy to the load. We define the quantities%
\begin{equation}
R_{12}=\frac{F_{1}}{F_{1}^{0}},T_{12}=\frac{F_{2}}{F_{1}^{0}},
\end{equation}%
as reflectance and transmittance, where $F_{i}$ is the final equilibrium
energy for each component and $F_{1}^{0}$ is the initial energy of the
active core. We remark that the total energy of the system in its final
equilibrium state is less than the initial energy of the active core due to
dissipation. In the limit that the outside radius of the load goes to
infinity, we obtain%
\begin{equation}
T_{12}=\frac{3E_{1}E_{2}}{(3E_{1}+E_{2})^{2}}.
\end{equation}%
Upon inserting an impedance matching host between the active core and the load,
the transmittance from the active core to the load becomes%
\begin{equation}
T_{13}=T_{12}T_{23}=\frac{9E_{1}E_{2}^{2}E_{3}}{%
(3E_{1}+E_{2})^{2}(3E_{2}+E_{3})^{2}}.
\end{equation}%
Maximizing $T_{13}$ over $E_{2}$ gives%
\begin{equation}
E_{2}=\sqrt{E_{1}E_{3}}.
\end{equation}%
We have recovered the results from above via an energy transfer point of
view in the limit when the size of the load goes to infinity. The main
difference between our case and light propagation case lies in that the energy is
not conserved in the former but is conserved in the latter case. It suggests
that energy conservation is not a key requirement in impedance matching
mechanisms.

\section{Conclusion}

In this work, we are interested in the work done by an active material on
the materials surrounding it. Specifically, we have an initially
nonequilibrium elastic system with all the energy stored in one part, and
the system is then allowed to relax. We look for ways to improve the
efficiency in transferring energy to other parts of the system at
equilibrium. To do so, we analyzed a composite system consisting an active
elastic material, a host, and a load. We found that in the cylindrical
geometry, the transferred energy from the active material to the load can be
maximized by tuning Young's modulus of the impedance matching host material.
The analysis was done using the linearized Mooney-Rivlin hyperelastic model
and assuming incompressibility of all components. We further assumed that
the system was overdamped and elastic wave propagation was not considered.
The active material located at the center of the host, when actuated, transfers
stored energy to the load through the host. We have considered two cases
where the outer boundary of the load is fixed and where it is free. Young's
modulus of the host material which maximizes the energy transfer is found to
be the geometric mean of the moduli of the active material and the load,
multiplied by a geometric factor which depends on the radii of the
components. In the limit when the size of the load goes to infinity, Young's
modulus for the host is simply the geometric mean of the moduli of the
active material and the load. This coincides with the classical result from
impedance matching in the case of light propagating through dielectric
media. Although the model is simplified with idealized geometry and is in the
small strain limit, we anticipate the results will help optimize the
performance of photomechanical materials by using an impedance matching host
between the active material and the load.

\section*{Acknowledgment}

This work was supported by the Office of Naval Research [ONRN00014-18-1-
2624]

\end{document}